# An Affordable Bio-Sensing and Activity Tagging Platform for HCI Research

Siddharth[1,2], Aashish Patel[1], Tzyy-Ping Jung[2], and Terrence J. Sejnowski[2,3]

[1] Department of Electrical and Computer Engineering, University of California, San Diego
[2] Institute for Neural Computation, University of California, San Diego
[3] The Computational Neurobiology Laboratory, Salk Institute

{ssiddhar@eng.ucsd.edu, anp054@eng.ucsd.edu, jung@sccn.ucsd.edu, terry@salk.edu}

**Abstract.** We present a novel multi-modal bio-sensing platform capable of integrating multiple data streams for use in real-time applications. The system is composed of a central compute module and a companion headset. The compute node collects, time-stamps and transmits the data while also providing an interface for a wide range of sensors including electroencephalogram, photoplethysmogram, electrocardiogram, and eye gaze among others. The companion headset contains the gaze tracking cameras. By integrating many of the measurements systems into an accessible package, we are able to explore previously unanswerable questions ranging from open-environment interactions to emotional-response studies. Though some of the integrated sensors are designed from the ground-up to fit into a compact form factor, we validate the accuracy of the sensors and find that they perform similarly to, and in some cases better than, alternatives.

## 1 Introduction

Electroencephalogram (EEG) systems have experienced a renewed interest by the research community for use in non-clinical studies. Though being deployed in large-scale studies, many of the advances have not been translated to substantial real-world applications. A major challenge is that the hardware and software typically used to make measurements limit their use to controlled environments. Additionally, the low spatial resolution of EEG itself limits the amount of usable information that can be extracted from noise in dynamic recording environments. Lastly, the absence of a method to automatically extract user-environment interactions for tagging with EEG data introduces an immense overhead to researchers - having to manually tag events or limit experimental design by requiring the subjects to provide information during the experiments.

Most of the EEG research from the past decades has been conducted under laboratory based controlled environments as opposed to practical daily-use applications. On the other hand, there are many fitness trackers available today capable of providing accurate heart-rate, blood pressure, galvanic skin response (GSR), steps taken etc. Under controlled laboratory conditions, EEG researchers have been able to control a quadcopter [1], control robots [2], control wheelchair to move around [3] etc. Unfortunately, research labs have been unable to show applications of EEG "into the wild" due to constraints imposed by the existing EEG decoders [17].

EEG research often studies event-related brain responses evoked or elicited by a visual or auditory stimulus. But, for real-world experiments with EEG, the stimulus onset is not measured or is ill-defined. A solution is to use saccadic eye movements and fixations as the time-locking mechanism for analyzing naturalistic visual stimuli [26, 27, 28]. Hence, we need to simultaneously record and synchronize EEG and eye-gaze data in real-world neuroimaging studies. For real-world experiments with EEG there is also a need to pinpoint the stimulus that is causing the changes in EEG. Hence, user's visual perspective is necessary to be recorded for EEG recordings in real-world experiments.

For the analysis of emotional responses, recent research [8, 10] (in NeuroCardiology) has shown that the heart also has a role to play in generation of emotions. This falsifies the wide ranging decades old belief that the brain is solely responsible for the generation and subsequent emotional feelings. But, there is no currently available system which can reliably sense and record EEG and electrocardiogram (ECG) together in a mobile environment. Furthermore, ECG complicates the experimental setup since subjects have to wear a belt or place several sensors on their chest. A workaround is to use photoplethsmogram (PPG) from commercially available devices that derive PPG from the wrist. But, such devices usually use low sampling rates to save battery power and hence can only measure heart rates, but not heart-rate variability (HRV) that is typically only estimated by commercial devices.

Addressing the above key limitations of existing systems, we present an affordable, wearable multi-modal bio-sensing platform that is capable of monitoring EEG, PPG, eye-gaze, and limb dynamics (Fig. 1). The platform also supports the addition of other biosensors including galvanic skin response (GSR) and lactate levels. Leveraging the capabilities of this system, a new breadth of applications can be explored that allow for better translations to impactful solutions.

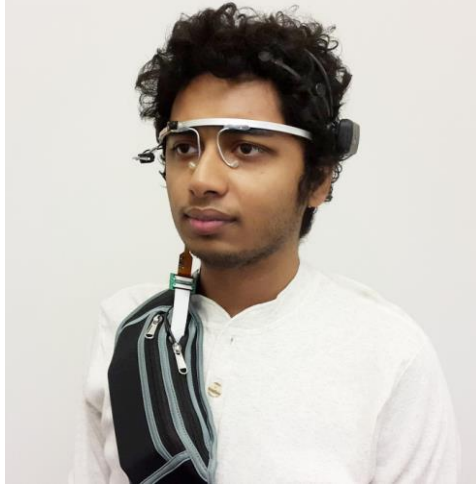

**Fig. 1.** Portable multi-modal bio-sensing platform paired with an Emotiv Epoc for EEG, PPG behind ear and eye-gaze collection.

## 2 System Overview

We use modular design to increase the flexibility and efficiency of multiple measurements of the multi-modal bio-sensing platform. Selecting a control board that is well supported by the open-source community and had capable expansion was a priority. To this end, this study has explored different solutions including the Arduino, Raspberry Pi, LeMaker Guitar, and other ARM-based embedded controllers. The hardware evaluation metric that determined viability was the ability for the systems to hit lower-bound frame-rates and collect data from multiple sensors in real-time using the Lab-Streaming Layer (LSL [23]). The last but one of the most important evaluation metrics was the expandability via general input/output or communication protocols. After evaluation of the different platforms, the Raspberry Pi 3 (RPi3) was identified as being the system that best balances cost, support, and capabilities. The sensors that were selected for preliminary use are explored in detail below.

### 2.1 Electroencephalogram (EEG)

Non-invasive EEG is used to collect neural signals from individuals. Any EEG system that is supported by LSL can be used in the proposed multi-modal bio-sensing framework. The Emotiv Epoc+ system is shown in Fig. 1 as it has a suitable tradeoff between ease-of-use and performance. The Epoc allows for wireless collection of data that can be time-stamped and synchronized in real-time by RPi 3. The sampling frequency of the system is on the lower end of new commercial systems at 128Hz, but has 14 channels (saline activated) and a gyroscope allowing for collection of cleaner signals. Independent Component Analysis (ICA) [5, 6, 21] is used in real time using ORICA [20, 24] toolbox in Matlab to separate the sources of EEG recordings in real-time and plot them. For each of the independent components, the scalp map is plotted in real-time to better depict the source localization. ICA is also used to remove EEG artifacts due to eye blinks, muscles and other movements.

### 2.2 Photoplethsmogram (PPG)

Due to the uncomfortable nature of existing heart-rate and heart-rate variability sensors, a new miniaturized PPG sensor (Fig. 2) was developed that magnetically clipped to the ear. The miniaturization was achieved by integrating a high-precision and high-sampling rate ADC to the sensor. Additionally, to eliminate noise, a third-order filter (bandpass 0.8 – 4 Hz) was also integrated on the board such that only the digitized and filtered signals are transmitted to the control board. To also account for motion artifacts in the heart-rate signals, a 3-axis accelerometer was integrated into the board. The two data streams, once collected by the core controller, are integrated using an adaptive noise cancellation algorithm (ANC) [7, 18] (Fig. 3). Addressing the discomfort and bulk associated with existing systems, the device was developed to be mountable to the ear-lobe using magnets [7]. Because the system is low-profile and capable of resting behind the ear [9, 16], more mobile studies can be conducted without the constrained natures of existing systems.

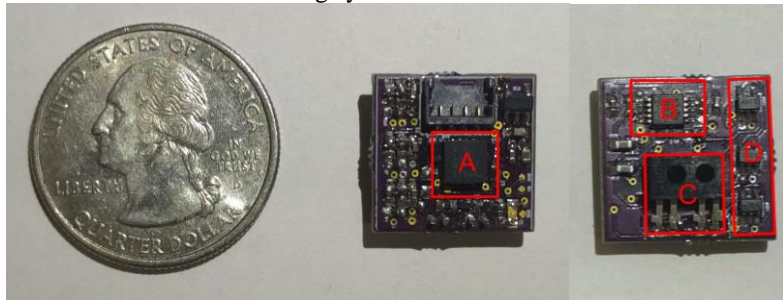

**Fig. 2.** Miniaturized PPG sensor with scale reference. (A) 3-axis accelerometer, (B) 100 Hz 12-bit ADC, (C) IR emitter and receiver, (D) third-order filter bank.

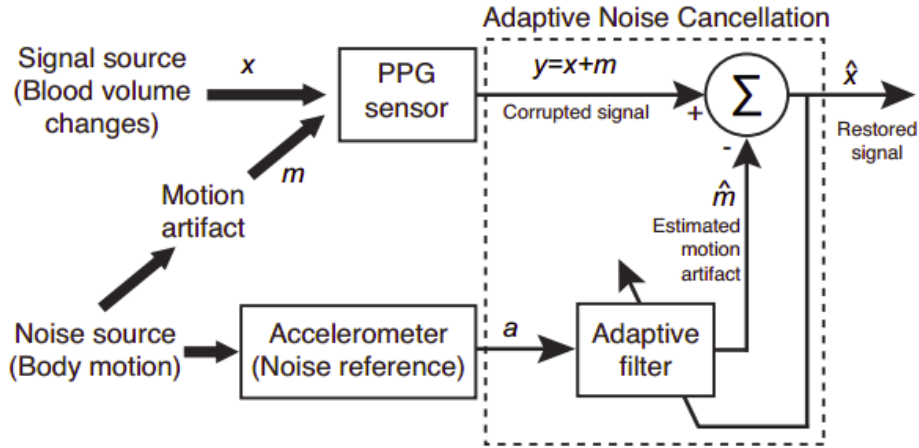

**Fig. 3.** Schematic overview of adaptive noise cancellation integration with PPG.

## 2.3 Eye Gaze

The next sensor of the multi-modal system is a pair of cameras. One camera, an IR emitting device, is capable of accurately capturing the pupil location. A pupil-centering algorithm is also integrated into the platform and is capable of maintaining the exact location even under perturbation. An algorithm developed by Pupil Labs [4] for pupil detection and eye-gaze calibration is utilized. Refer to the results section for quantification of tracking accuracy.

The second integrated camera in the system is a world-view camera. The camera provides a wide-angle view of what the wearer is seeing. While being small and integrated into the headset, the camera itself is a standard easily-accessible module. With the information that is retrievable from both the pupil and the world cameras, it is possible to retrospectively reconstruct the full-view that the user was observing. The primary problem that stems from this type of mass video collection is that the amount of data that must be manually labelled is enormous. There are machine-learning tools that are capable of labelling video post-hoc, but limit the types of experiments that can be performed. To create a truly portable system, the system's video can be streamed to a computer and processed using deep-learning libraries such as You Only Look Once (YOLO) [19] that are capable of labelling 20 objects in real-time (trained on Pascal VOC [20] dataset). By labelling exactly what the user is looking at and allowing labelled data to be accessible during the experiment, the experimental rigidity can be relaxed allowing for more natural free-flowing behavior to be measured with minimally intrusive cues (Fig. 4).

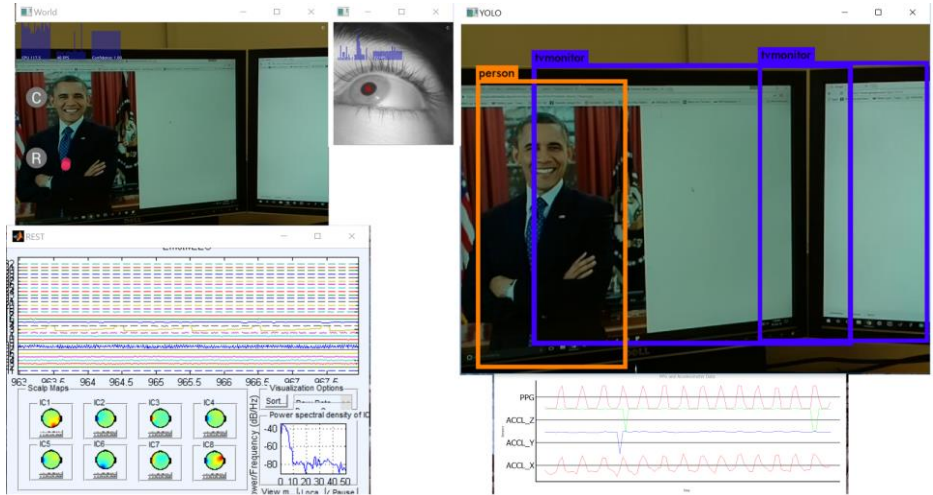

**Fig. 4.** Pupil and world views from companion headset device (top-left). Deep-learning package used to classify objects in real-time (top-right). EEG with real-time ICA and PPG signals capture (bottom panels).

### 2.4 Galvanic Skin Response (GSR)

The final sensor considered for addition to our multi-modal setup is a galvanic skin response sensor. GSR specifically allows for the measurement of arousal through the measurement of the skin's impedance. This sensor is unique in that the efficacy of a third party commercial product being integrated into this research platform needed to be explored. The GSR sensor that was selected for use was the Microsoft Band 2 [14].

## 3 Evaluation

The proposed device addresses many of the limitations of existing systems while providing the measurement capabilities in a form-factor that is convenient for both researchers and subjects. To evaluate the efficacy of the system, the individual components that were created in this study were evaluated. In particular, the evaluations of the Emotiv Epoc and Microsoft Band are not explicitly evaluated in this review. The novel PPG and eye-gaze tracking systems will be evaluated for effectiveness in their respective areas.

### 3.1 PPG Evaluation

To quantify the performance of the miniaturized PPG sensor, different scenarios are considered that are representative of real-world uses. The baseline system for comparison is an EEG/ECG collection system from the Institute of Neural Engineering of Tsinghua University, Beijing, China. It is capable of measuring EEG/ECG at

1,000 Hz. Because the reference system takes measurements from electrodes placed near the heart, the artifacts introduced from movements or other physiological responses are minimized. Simultaneously while taking measurements from the reference system, the PPG is collecting the ECG signal from the user's ear at a rate of 100 Hz. As both systems can be connected in parallel, they are synchronized using the lab-streaming layer [23] and analyzed post-hoc.

The first experiment was a resting scenario - the user remained seated for a fixed period of two minutes. For the PPG sensor, the data were compared to the reference with and without the adaptive noise cancellation filter. The second experiment was an active scenario where the user was instructed to walk in-place at a normal pace to simulate an active walking scenario. Again the data after using adaptive noise cancellation was compared against the standard raw PPG signal.

A peak detection algorithm [25] using minimum distance to next peak as one of the parameters to eliminate false peaks was used to calculate Heart Rate (HR) from ECG and PPG Data. Fifteen-second trails were used to calculate the HR using the peak-detection algorithm. Figures 5 & 6 show the normalized errors, the ratio of the difference in HR between PPG and ECG-based methods divided by the mean HR obtained by PPG. A perfect HR estimation should result in 0%. Examining the results from the reference signal, the ANC enabled, and ANC disabled signals, it is clear that the ANC enabled signals have the least amount of noise and most closely match the reference signal. For resting, the ANC-disabled signals were nearly undistinguishable from the ANC-enabled signals (Fig. 5). It is in active environments that having the ANC filtering provide a marked improvement in noise rejection (Fig. 6).

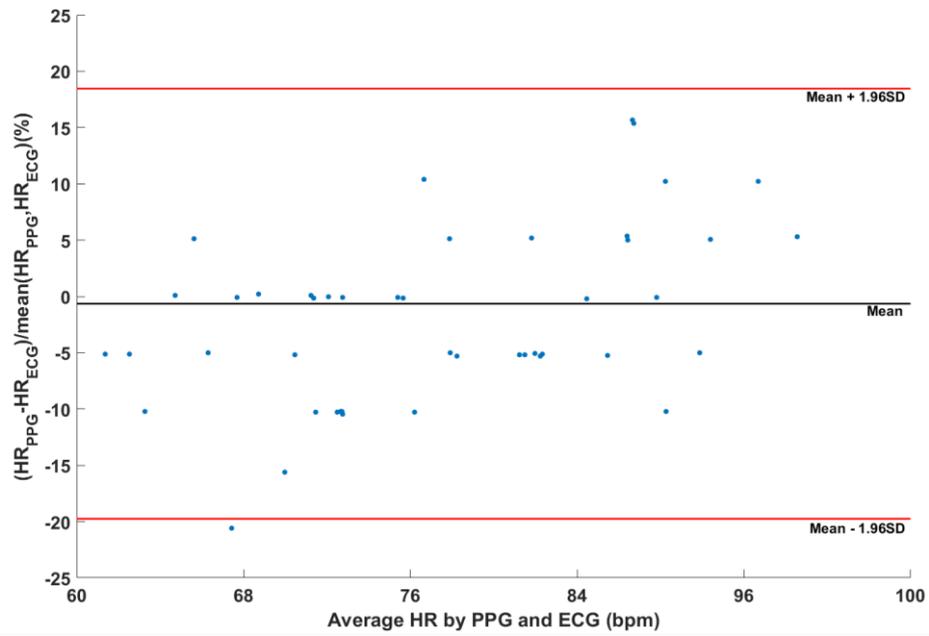
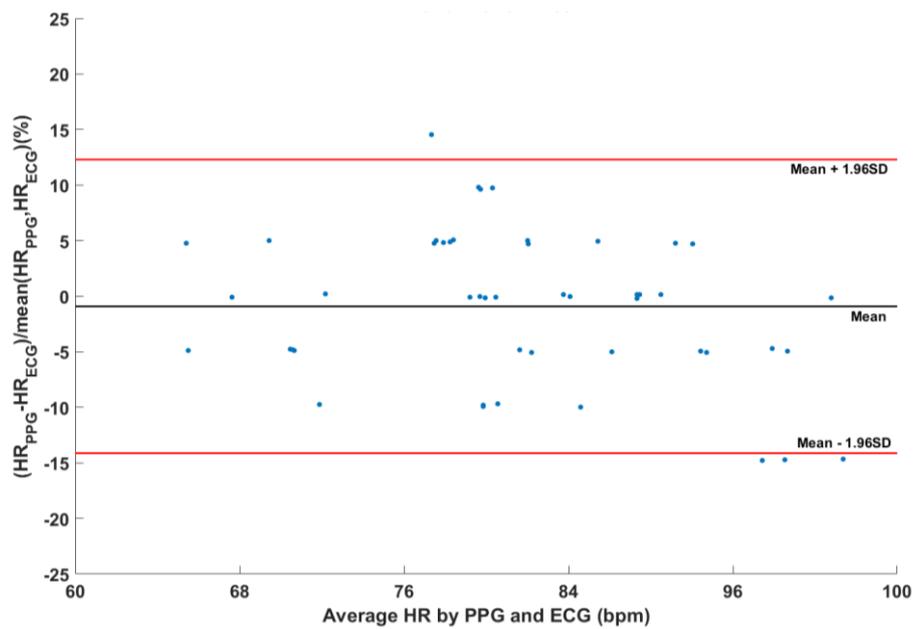

**Fig. 5.** Bland-Altman plot comparing the measured PPG signal to a reference while at rest (top). Similarly comparing the measured PPG signal using an adaptive noise cancellation filter to reference while at rest (bottom).

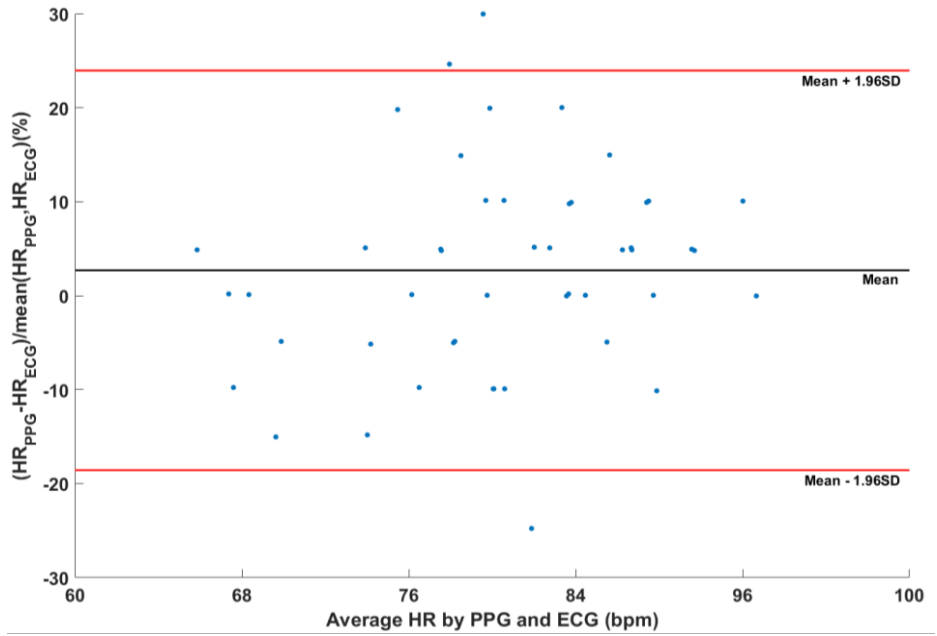

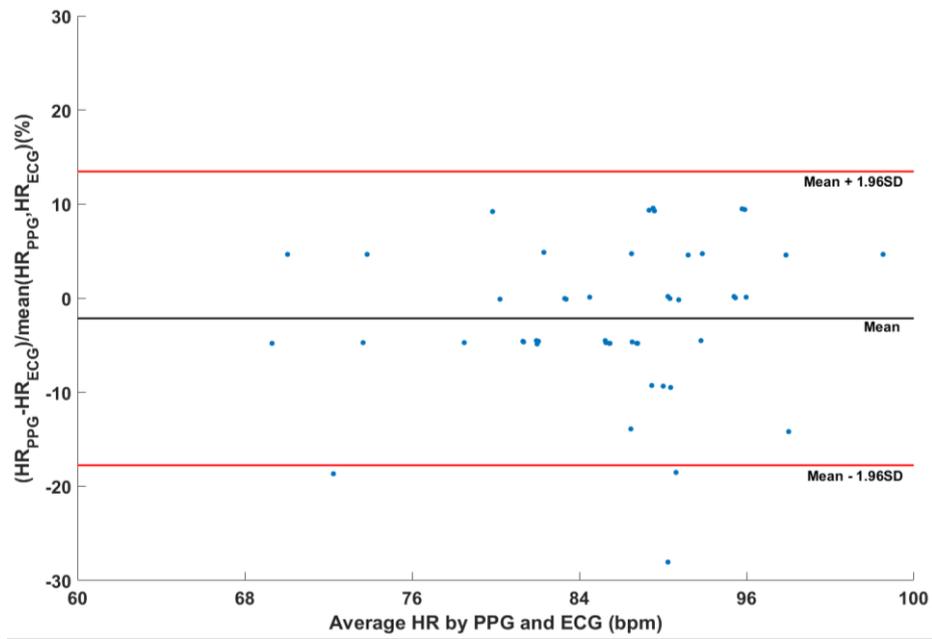

**Fig. 6.** Bland-Altman plot comparing the measured PPG signal while walking (top). Similarly comparing the measured PPG signal using an adaptive noise cancellation filter while walking (bottom).

### 3.2 Eye Gaze Evaluation

The performance of the paired pupil- and world- view cameras was evaluated using a structured visual task to measure precision and accuracy during use. The user sat 2-2.5 feet away from a computer monitor such that the world camera was >90% of the camera view was composed of the task screen. Both cameras were streamed at 30fps. For the first task, the participant was instructed to fix their head movement and only move their eyes to gaze at static targets that appeared on the screen. A calibration step where 9 targets appeared in a regular fashion on the screen calibrated the user's gaze marker. Immediately following the calibration process, a series of 20 unique targets are collected in distributed manner across the full screen accounting for the majority of the field of view. This process was followed by a period of 30 seconds of rest where the user was asked to move their head around without removing the headset. This action was designed to simulate the active movement scenarios when wearing the headset. Next, the participant is instructed to return to a preferred position and maintain head position. Twenty new unique points are shown on the screen to measure the precision and accuracy of the eye-tracking system after active use. This process was repeated for a total of three trials per subject.

Examining the results for the patients, we are able to observe that the accuracy and precision of the eye gaze setup does not drift significantly from the expected output. The accuracy is measured as the average angular offset (distance in degrees of the visual angle) between fixation locations and the corresponding fixation targets (Fig. 7). The precision is measured as the root-mean-square of the angular distance (degree of visual angle) between successive samples during a fixation (Fig 8.). Compared to literature, the gaze accuracy drift of 0.42 degrees is significantly less than the 1-2 degree drift found in commercial systems [11, 12]. The precision, on the other hand, experiences only a 0.2 degree shift post movement, indicating a minimal angular distance shift.

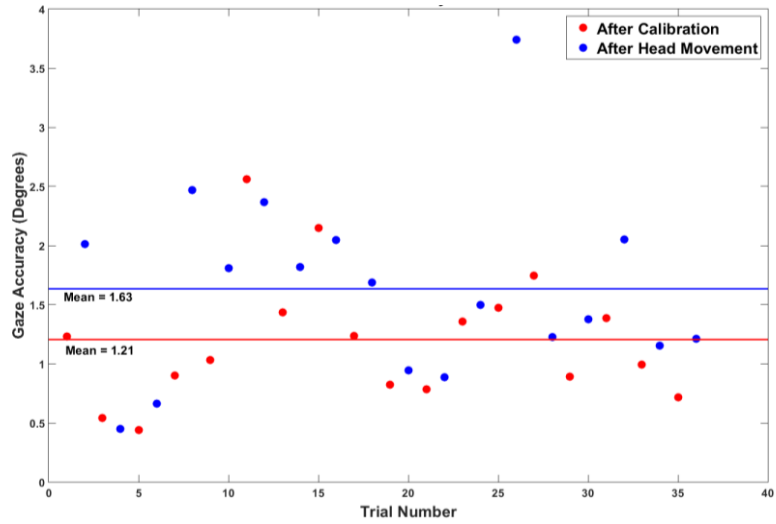

**Fig. 7.** Gaze accuracy analysis comparing the mean after calibration (red) and after 30 seconds of dynamic head movement to simulate active conditions (blue).

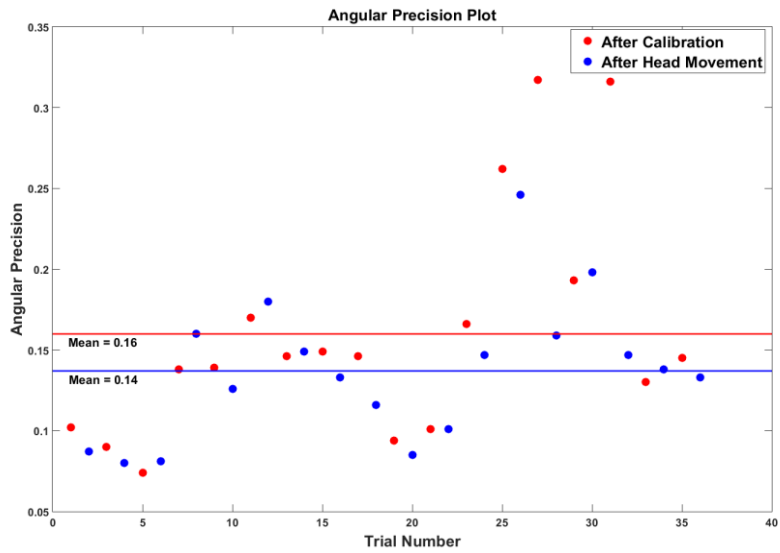

**Fig. 8.** Angular precision analysis comparing the mean after calibration (red) and after 30 seconds of dynamic head movement to simulate active conditions (blue).

# 4    Conclusion

There are numerous sensors capable of measuring useful metrics for human behavior and interactions, however, limitations in the collection hardware and software hinder their use in experiments spanning multiple modalities. By developing a low-cost, portable, multi-modal bio-sensing platform that is capable of interfacing with numerous different sensors, we are able to explore richer experimental questions that have previously been unable to be accessed due to the constrained nature of the measurement hardware. In particular, the modular nature of the control board, interface software, and headset, time can be better spent looking for novel research insights rather than wrangling devices and software packages from different manufacturers.